\documentclass[acmsmall, nonacm]{acmart}

\usepackage[english]{babel}
\usepackage{array}
\usepackage{fancyvrb}
\usepackage{hyperref}
\usepackage{relsize}
\usepackage{textcomp}
\usepackage{xcolor}
\usepackage{xspace}
\usepackage{booktabs}
\usepackage{listings}
\usepackage{smartdiagram}
\usepackage{scalefnt}
\usepackage[nolist]{acronym}

\usetikzlibrary{shapes}
\usetikzlibrary{intersections}
\usetikzlibrary{calc}
\usetikzlibrary{optics}
\usetikzlibrary{arrows}
\usetikzlibrary{positioning}
\usetikzlibrary{decorations}
\usetikzlibrary{angles}
\usetikzlibrary{quotes}
\usetikzlibrary{arrows.meta,shapes.geometric,decorations.pathreplacing}

\definecolor{backcolour}{rgb}{0.95,0.95,0.92}

\lstdefinestyle{codestyle}{
    backgroundcolor=\color{backcolour},
    basicstyle=\ttfamily\footnotesize,
    breakatwhitespace=false,
    breaklines=true,
    captionpos=b,
    keepspaces=true,
    numbersep=5pt,
    showspaces=false,
    showstringspaces=false,
    showtabs=false,
    tabsize=2
}

\addto\extrasenglish{

}

\usepackage{siunitx}
\newcolumntype{L}[1]{>{\raggedright\let\newline\\\arraybackslash\hspace{0pt}}m{#1}}
\newcolumntype{C}[1]{>{\centering\let\newline\\\arraybackslash\hspace{0pt}}m{#1}}

\definecolor{olivegreen}{rgb}{0.0, 0.6, 0.0}
\definecolor{darkred}{rgb}{0.7, 0.0, 0.0}
\definecolor{orangeyellow}{rgb}{0.9, 0.7, 0.2}
\newcommand{\red}[1]{\textcolor{darkred}{#1}}
\newcommand{\orange}[1]{\textcolor{orangeyellow}{#1}}
\newcommand{\green}[1]{\textcolor{olivegreen}{#1}}

\newcommand{\gh}{GitHub\xspace}
\newcommand{\gl}{GitLab\xspace}
\newcommand{\git}{Git\xspace}
\newcommand{\docx}{\textit{.docx}\xspace}
\newcommand{\wftw}{Word for the Web\xspace}
\newcommand{\magit}{markup\,+\,\git}

\newcommand{\Rplus}{\protect\hspace{-.1em}\protect\raisebox{.35ex}{\smaller{\smaller\textbf{+}}}}
\newcommand{\Cpp}{\mbox{C\Rplus\Rplus}\xspace}

\AtBeginDocument{%
\providecommand\BibTeX{{%
\normalfont B\kern-0.5em{\scshape i\kern-0.25em b}\kern-0.8em\TeX}}}

\copyrightyear{}
\acmYear{}
\setcopyright{none}

\acmBooktitle{}
\acmPrice{}
\acmISBN{}
\acmDOI{}

\acmSubmissionID{}

\begin{document}

\title{Large-Scale Collaborative Writing: Technical Challenges and Recommendations}


\author{Markus Hofbauer}
\authornote{Both authors contributed equally to this research.}
\email{markus.hofbauer@tum.de}
\orcid{0000-0002-8167-5485}
\affiliation{%
    \institution{Technical University of Munich}
    \department[2]{School of Computation, Information and Technology}
    \department[1]{Department of Computer Engineering}
    \department[0]{Chair of Media Technology}
    \department[0]{Munich Institute of Robotics and Machine Intelligence (MIRMI)}
    \city{Munich}
    \state{Bavaria}
    \country{Germany}
    \and
    \institution{Luminar Technologies, Inc.}
    \city{Munich}
    \state{Bavaria}
    \country{Germany}
}

\author{Christoph Bachhuber}
\authornotemark[1]
\email{christoph.bachhuber@luminartech.com}
\orcid{0000-0002-9808-0162}
\affiliation{%
    \institution{Luminar Technologies, Inc.}
    \city{Munich}
    \state{Bavaria}
    \country{Germany}
}

\author{Christopher Kuhn}
\email{christopher.kuhn@tum.de}
\orcid{0000-0002-0107-0516}
\affiliation{%
    \institution{Technical University of Munich}
    \department[2]{School of Computation, Information and Technology}
    \department[1]{Department of Computer Engineering}
    \department[0]{Chair of Media Technology}
    \department[0]{Munich Institute of Robotics and Machine Intelligence (MIRMI)}
    \city{Munich}
    \state{Bavaria}
    \country{Germany}
}

\author{Sebastian Schwarz}
\email{sebastian.schwarz@nokia.com}
\affiliation{%
    \institution{Nokia Technologies}
    \city{Munich}
    \state{Bavaria}
    \country{Germany}
}

\author{Bart Kroon}
\email{bart.kroon@philips.com}
\affiliation{%
    \institution{Philips Research Eindhoven}
    \city{Munich}
    \state{Bavaria}
    \country{Germany}
}

\author{Eckehard Steinbach}
\email{eckehard.steinbach@tum.de}
\orcid{0000-0001-8853-2703}
\affiliation{%
    \institution{Technical University of Munich}
    \department[2]{School of Computation, Information and Technology}
    \department[1]{Department of Computer Engineering}
    \department[0]{Chair of Media Technology}
    \department[0]{Munich Institute of Robotics and Machine Intelligence (MIRMI)}
    \city{Munich}
    \state{Bavaria}
    \country{Germany}
}

\begin{acronym}
    \acro{ISO}{International Standardization Organization}
    \acro{CI}{continuous integration}
    \acro{VCS}{version control system}
    \acro{WYSIWYG}{what-you-see-is-what-you-get}
    \acro{URL}{Uniform Resource Locator}
    \acro{PR}{Pull Request}
    \acro{MR}{Merge Request}
    \acro{OS}{operating system}
    \acro{CSS}{cascading style sheets}
    \acro{IP}{intellectual property}
    \acro{MPEG}{Motion Picture Experts Group}
\end{acronym}


\begin{abstract}
    Collaborative writing is essential for teams that create documents together.
    Creating documents in large-scale collaborations is a challenging task that requires an efficient workflow.
    The design of such a workflow has received comparatively little attention.
    Conventional solutions such as working on a single Microsoft Word document or a shared online document are still widely used.
    In this paper, we propose a new workflow consisting of a combination of the lightweight markup language AsciiDoc together with the state-of-the-art \acl{VCS} \git.
    The proposed process makes use of well-established workflows in the field of software development that have grown over decades.
    We present a detailed comparison of the proposed \magit workflow to Word and \wftw as the most prominent examples for conventional approaches.
    We argue that the proposed approach provides significant benefits regarding scalability, flexibility, and structuring of most collaborative writing tasks, both in academia and industry.
\end{abstract}


\maketitle

\section{Introduction}

Documentation of a project is vital for successfully sharing knowledge in both academia and industry~\cite{Halsey2019}.
In collaborative writing, multiple individuals need to access the same content, often at the same time, which makes documentation a complex task.
Currently, many individuals and organizations simply exchange files such as Microsoft Word documents for collaborative editing of documents.
A reason for this workflow can be that the first draft of the document was created as such a file due to simplicity, or that this process is the traditional way at a workplace.
As soon as more people are getting involved, the single document is then shared with them.
This practice is not only the case for small student teams working on a short-lived essay, but widespread throughout all levels of industry and academia for various document types.
To give a prominent example, the \ac{ISO} requires committees to submit standard drafts as Word files~\cite{iso_word_template}.
These committees, which can consist of dozens of experts, therefore need to work on shared Word documents with potentially hundreds of pages.
Microsoft Word has not been designed for distributed, large-scale collaboration on a set of files.
Consequently, the processes required to collaborate this way are inefficient, not scalable, and often lack structure.

In contrast, \acp{VCS} such as SVN, Mercurial, and \git~\cite{chacon_pro_2014} have been created for version control and collaboration on plain text source code in software projects.
Plain text markup languages such as Markdown and AsciiDoc are therefore suitable for \acp{VCS} such as \git and enable collaboration on proven platforms such as \gh or \gl.

In this paper, we propose a new workflow for collaborative writing that uses the benefits from systems used for software development.
We propose a combination of a plain text based markup language such as AsciiDoc, a \ac{VCS} such as \git, and a collaboration platform such as \gh or \gl.
For large scale collaborative writing, we argue that this approach is a superior alternative to Word or comparable office suites based on binary document types as well as shared online platforms such as \wftw.

In summary, we make the following contributions:

\begin{itemize}
    \item We discuss existing tools for collaborative writing.
    \item We propose a framework for efficient collaborative writing that makes use of the benefits of modern \acp{VCS}, markup languages, and collaboration platforms.
    \item We compare the proposed approach to Word and \wftw as representatives for established tools.
\end{itemize}

The rest of this paper is organized as follows.
In \autoref{sec:collaboration}, we summarize existing methods in the field of version control and collaborative writing.
Next, we introduce the proposed framework for efficiently managing a large-scale collaboration such as an \ac{ISO} standardization process in \autoref{sec:process}.
In \autoref{sec:comparison}, we present a detailed comparison of Microsoft Word and \wftw to the proposed approach.
\autoref{sec:conclusion} concludes the paper.

\section{Collaboration Theory and Tools}
\label{sec:collaboration}

This section provides an overview of theory and tools for collaborative work.
First, we give a general overview of the challenges of collaboration.
Then, we summarize \ac{WYSIWYG} tools such as Microsoft Word, LibreOffice Writer, or Google Docs.
Further, we discuss tools which distinguish plain text input from the rendered output document such as Markdown, AsciiDoc, and LaTeX.

\subsection{Collaboration Theory}

The benefits of collaborative writing are widely recognized~\cite{storch2019collaborative}.
Computer-supported collaborative work has been studied since the introduction of computers~\cite{grudin1994computer}.
Schutzler~et~al.~\cite{schuetzler2019learning} have further shown the benefit of using collaboration via \git for teaching and learning new concepts.
The field of collaboration engineering focuses on the design of efficient collaboration processes that can be repeatedly applied by non-expert practitioners~\cite{de2019program}.
In this paper, we do not design a new process or contribute to the theory of collaboration, but focus on which tools are most suitable for the established process of collaborative writing.
An important component when designing tools for this task is to consider the human factor in collaboration.
Some works highlight the importance of not dividing human attention too much, but to ensure that each collaborator can focus on a single task at hand~\cite{arias2000transcending}.
The proposed workflow allows each contributor to be assigned specific issues and only work on their own version of the document.
Works such as~\cite{edwards2004implementing} discuss the significant challenges of virtual teamwork, regardless of which tools are used.
While Muri\'{c}~et~al.~\cite{muric2019collaboration} show that the first few additional collaborators increase the productivity of each individual, they also demonstrate that the productivity decreases for larger groups due to communication and process overhead.
The question of whether technologies such as \git and AsciiDoc are the right approach for the general task of collaborative writing is a typical example of task-technology fit~\cite{goodhue2006task}.
We argue that to overcome the challenges of virtual teamwork, using tools developed over decades of successful software engineering collaboration is the most promising direction.

\subsection{Local \ac{WYSIWYG} with Microsoft Word}

Microsoft Word is a proprietary \ac{WYSIWYG} application, first published in 1983 and continuously improved until its latest version Word 2021.
Word is a suitable representative since Microsoft's Office suite, with Word being one of the major tools, had an \SI{87.5}{\percent} market share in 2018~\cite{schwartz_microsoft_2020}.
It is available for all major desktop and mobile \acp{OS}, with the exception of Linux-based \acp{OS}.
The latest version provides a comprehensive featureset, covering needs from font modification, layout options, referencing, bibliography management, to complex mathematical equations.
It is used as a universal tool for a great variety of document types, e.g. school essays, books, scientific articles, patents, international standards, meeting reports, and quick notes.

Word has the following relevant characteristics:

\subsubsection{Beginner Friendliness}

The graphical user interface allows for straightforward discovery of the available features.
Further, it provides templates for typical documents such as curricula vitae, calendars, and letters.
Such features allow for less experienced users to quickly achieve results.
This beginner friendliness, combined with the close resemblance to sheets of paper, explains why Word is often taught as a first tool for creating documents.

\subsubsection{Collaboration Tools}

The user can toggle a review mode in which all changes are tracked.
In this mode, the user can add comments on the selected text.
Collaborators can respond to comments, but given the simple design of Word comments, extensive, complex discussions cannot take place there.
Once comments are resolved, a list of all resolved comments can be displayed in a dedicated view.
For comparing two versions of the same document, Word provides a difference view.

\subsubsection{File Structure}

One document is a single \docx file.
It contains all text and images in a proprietary format which does not interface with other applications.
Applications such as LibreOffice Writer or Pandoc can parse and create \docx files, but interfacing is imperfect and can result in layout or content errors.
The single-file approach renders navigating long Word documents cumbersome.
Word offers a navigation pane with links to sections to alleviate this issue.

\subsubsection{Correctness}

Autocorrection detects spelling as well as grammatical errors and suggests improvements, which informs the user and enhances the quality of the text created.

\subsection{Online \ac{WYSIWYG}}
\label{ssec:word-online}

Web-based \ac{WYSIWYG} tools such as \wftw and Google Docs resolve many of the issues of collaborating with manually shared files.
Such tools do not require collaborators to download a file and open a standalone application.
Instead, the file resides on a shared (cloud-)storage, to which all collaborators must have access.
Online \ac{WYSIWYG} tools are web applications, requiring only a modern browser for editing the document.
As a representative, we introduce \wftw in more detail in the following.

When \docx files are stored in online storage such as OneDrive or SharePoint, they can be edited with \wftw as well as the standalone Word application.
\wftw mirrors most of the Word standalone application's user experience, with some notable differences.
\wftw lacks equations, advanced table tools, SmartArt, charts, signature, drawing and design dialogs, bibliography and captions, and many more.

By itself, each of these differences is not a significant issue.
However, given the large number of discrepancies, the lack of tools in \wftw will force users to go back to the standalone application for various tasks.
\wftw offers several features which are not found in the standalone application.
First, it allows for live collaborative editing in the same document.
When editing a shared document with the web and the desktop versions simultaneously, merge conflicts arise immediately after edits from different applications.
Second, a reuse-files tool allows searching through existing files in OneDrive/SharePoint to avoid duplicate text or writing.
Finally, all files are versioned automatically.
\wftw automatically saves a file upon new changes and offers a file version history.
However, the user has no control over when versions are created.
The versions can only be identified by date and time.
There is no possibility for annotating versions with meaningful messages.

Given the differences between the web and standalone applications, collaboration using both application types requires switching between the applications to get access to all features.
Additionally, manual resolution of unnecessary merge conflicts created by the tool is required.

\subsection{Version Control}

Engineering leaders at Google state: "\textit{Perhaps no software engineering tool is quite as universally adopted throughout the industry as version control. One can hardly imagine any software organization larger than a few people that doesn`t rely on a formal Version Control System (VCS) to manage its source code and coordinate activities between engineers.}"~\cite{winters_software_2020}

Fundamentally, software engineering is a collaborative project on a set of documents, the source code files.
Thus, adopting version control for other kinds of collaborative document editing might entail similar benefits as seen in the software industry.

Version control has seen tremendous adoption rates in software engineering over the past two decades, most notably \git~\cite{german_continuously_2016}.
Additionally, other fields such as scientific research adopt version control for its benefits in transparency, collaboration features, reproducibility, and time savings \cite{ram_git_2013, lowndes_our_2017}.

\subsubsection{\git Workflow}
\label{sec:git}

The open source tool \git~\cite{chacon_pro_2014} is the de facto standard for file version control in open source~\cite{openhub_git_2021} and commercial projects.
\git supports distributed collaboration and is highly secure and efficient.

Branches are an integral part of a typical \git workflow.
If a contributor wants to change something, a branch is created based on the latest accepted version of the files which is called \textit{trunk}, \textit{head}, \textit{master}, or \textit{main}.
Next, the contributor can perform all changes on that branch in their isolated environment.
Other collaborators and the \textit{head} are not affected by those actions.
Once all changes are implemented, the contributor requests to merge the changes back into the main branch.
Depending on the platform, such requests are either called \ac{MR} or \ac{PR}.
At this point, \ac{PR} reviews, further detailed in the next section, come into play.
When the changes are merged, the branch is discarded.
New branches can then be created for the next work packages.
As the process of merging is a core feature of \git, this is highly optimized and can be performed automatically in many cases.
Other workflows with non-main branches merging into each other are also possible.

Further, \git is a decentral, distributed \ac{VCS} in which each collaborator has their own complete set of files.
No connection to a central server is required while working on the files.
To ease coordination, projects usually choose a central hosting server such as \gh or \gl as the root repository containing the latest accepted changes.

Versioning in \git is done via commits.
From a user perspective, a commit contains a change to one or more files.
Contributors create commits when they think that they have completed a unit of their work package.
A commit contains a timestamp, commit message, unique commit hash, author, and additional metadata.
A \git history is simply a sequence of commits on one or more branches.
Given complete control over commit creation, teams usually create expressive \git histories that can be navigated efficiently.
Modern editors such as VS Code~\cite{vs_code} provide a deep integration of \git features.
For instance, the latest commit including some meta information that affected a certain line of text can be displayed.
This can drastically reduce the time of error tracing.

\subsubsection{\gh}
\label{sec:github}

Depending on the project needs, a \git project can be hosted publicly or on a private/company server.
\gh is a popular service for hosting document collections in projects, which are separate \git repositories.
Each project can be managed individually, e.g. in terms of user access level.
In \cite{longo_use_2015}, the authors already analyzed that \gh is used beyond classical software engineering projects and a well-suited solution for open collaboration on text documents.
Here, we summarize the key features of \gh that are most important for collaborative document creation.

One of the major features of \gh are \acfp{PR}.
A collaborator opens a \ac{PR} to merge a work branch into the repository's main branch.
PRs allow structured discussions that are permanently visible.

Reviewers can be explicitly assigned to request their review.
Other interested contributors can also review the \ac{PR}.
Reviewers see the changes of one work branch in isolation, which allows them to focus on the task at hand.
If required, they can access the entire document at the latest change for more context.
Multiple \acp{PR} may take place in parallel, but they are isolated from each other.
Each discussion is focused on one topic.
Difference views are independent of each other.
Avoiding double work might require some coordination through the management tools presented in the next subsubsection.
When a reviewer comments on a line, a discussion thread starts below that comment.
When the collaborators involved in the discussion agree on a solution, the comment is marked as resolved.
The solution can be a modification to the proposed change, a follow-up task, or an agreement in the discussion.
All comments and the related \acp{PR} remain visible and easily accessible through permanent \acp{URL}.
This provides traceability, which is required in some contexts and useful in most projects in order to see the history and reasons for past decisions.
If all comments of a reviewer are resolved, the reviewer accepts the changes and marks the \ac{PR} as \textit{approved}.
Different repository policies are possible, such as requiring one or two approvals before a proposed change can be merged into the main branch.

\gh provides management tools to streamline collaboration.
Workloads are organized using \textit{issues}.
An issue can be a discussion thread, a proposal for new content or a problem in the existing document, for example.
When the related work package is clear, an assignee is appointed and starts working on the issue.
In the process, the assignee can open multiple \acp{PR} and link to them, again providing traceability.
One or more labels can be assigned to issues and \acp{PR} for the organization of workloads into topics.
Labels can be used as filters.
Finally, milestones are used to group issues.

\subsection{Lightweight Markup Languages}

In markup languages, a plain text document can be annotated with elements that are syntactically different from the text.
When the plain text document is processed for display, these elements act as instructions to format the text rather than a direct visualization.
Examples for markup languages are HTML, \LaTeX, Markdown, and AsciiDoc.
As an example, we show the following Markdown plain text:

\begin{lstlisting}
# First Heading
## Second Heading
Markdown can do **bold** and *italic* text,
see [Wikipedia](www.wikipedia.de).
\end{lstlisting}

This text can be, depending on the Markdown toolchain, rendered to

\noindent
\fbox{\begin{minipage}{\columnwidth}
        \begin{flushleft}
            {\LARGE \underline{1.
                    First Heading}}

            {\large 1.2 Second Heading}

            Markdown can do \textbf{bold} and \textit{italic} text, see \href{www.wikipedia.de}{\color{blue}{Wikipedia}}.
        \end{flushleft}
    \end{minipage}}

Symbols such as \#, **, * and patterns such as []() are markup instructions that control formatting and are not shown in the displayed text.

With formatting instructions that are deliberately easy to use and memorable, markup languages targeting human usage such as Markdown and AsciiDoc create an efficient framework for text processing~\cite{thomas2019pragmatic}.
Some markup languages have fewer formatting options compared to Word, but they still suffice for most documents and can even prevent antipatterns, such as putting overly complex layout elements into table cells.
In particular, AsciiDoc seems to have found a good balance between simplicity of usage and a sufficient feature set for many use cases.
AsciiDoc's semantics are similar to Markdown and offer some more built-in features such as \LaTeX-based equations.
AsciiDoc has been used to write books~\cite{chacon_pro_2014, ramalho_fluent_2015}, extensive software documentation such as the Khronos\textsuperscript{\textregistered} Vulkan\textsuperscript{\textregistered} API~\cite{leech_khronos_2021}, and is used in some open source projects for documentation.
Further, Marquardson~et~al.~\cite{marquardson_learning_2019} used AsciiDoc together with \gh in education, where students should collaboratively create a tutorial for a certain topic.

Markup languages are furthermore easily extensible, e.g., via \ac{CSS} to support special use cases.
A prominent example is the Markdown-based bitstream specification~\cite{google_draco_bitstream_2021} of Google's Draco 3D compression library~\cite{google_draco_2021}.
For the table-based syntax elements of Draco's specification, Google created a \ac{CSS} element to display simple plain text bitstream specifications as tables.

Given that markup languages are plain text documents, any editor can be used to process them.
There are widely used editors such as VS Code~\cite{vs_code}, which have built-in support for previewing the rendered document while editing it.
Collaboration platforms such as \gh or \gl provide a preview of the rendered markup directly inside the browser.

\subsection{Collaboration Scenarios}

Next, we discuss common collaboration scenarios and the conventional workflow used in those collaborations.

\subsubsection{Standards and Intellectual Property Documents}
\label{sec:standards-workflow}

\begin{figure}[ht!]
    \centering
    \resizebox {.85\columnwidth} {!}{\scalefont{0.9}
\begin{tikzpicture}

    \tikzset{block/.style= {draw, rectangle, align=center,minimum width=1cm,minimum height=0.5cm}}

    \node[](anchor){};

    \node [block, text width = .9\columnwidth,fill=gray,fill opacity = 0.1, text opacity = 1, below = 0cm of anchor]  (get_recent) {
            \textbf{1. Get most recent version}
            \vspace{-.8em}
            \begin{itemize}
                \setlength\itemsep{-.4em}
                \item Requires access via messenger or platform
                \item Requires verification that correct file is selected
            \end{itemize}};

    \node [block, text width = .9\columnwidth,fill=gray,fill opacity = 0.1, text opacity = 1, below = 0.5cm of get_recent]  (plan_changes) {
            \textbf{2. Plan changes}
            \vspace{-.8em}
            \begin{itemize}
                \setlength\itemsep{-.4em}
                \item Inform collaborators about planned changes
                \item Ensure no parallel changes to avoid conflicts
            \end{itemize}};

    \node [block, text width = .9\columnwidth,fill=gray,fill opacity = 0.1, text opacity = 1, below = 0.5cm of plan_changes]  (implement_changes) {
            \textbf{3. Implement changes}
            \vspace{-.8em}
            \begin{itemize}
                \setlength\itemsep{-.4em}
                \item Enable review mode to track changes
                \item Perform changes in document and save to file
            \end{itemize}};

    \node [block, text width = .9\columnwidth,fill=gray,fill opacity = 0.1, text opacity = 1, below = 0.5cm of implement_changes]  (review) {
            \textbf{4. Review process}
            \vspace{-.8em}
            \begin{itemize}
                \setlength\itemsep{-.4em}
                \item Send file to all reviewing collaborators
                \item Requires coordination to merge all comments
            \end{itemize}};

    \node [block, text width = .9\columnwidth,fill=gray,fill opacity = 0.1, text opacity = 1, below = 0.5cm of review]  (finalize) {
            \textbf{5. Finalize update}
            \vspace{-.8em}
            \begin{itemize}
                \setlength\itemsep{-.4em}
                \item Document update, e.g. by changing file name
                \item Share agreed version with all collaborators
            \end{itemize}};

    \path[draw, -{Latex[length=2.5mm,width=1.5mm]}]
    (get_recent.south) edge (plan_changes.north)
    (plan_changes.south) edge (implement_changes.north)
    (implement_changes.south) edge (review.north)
    (review.south) to (finalize.north);
\end{tikzpicture}}
    \caption{Overview of the conventional workflow used for applications such as \ac{ISO} standards.}
    \label{fig:workflow_sota}
\end{figure}
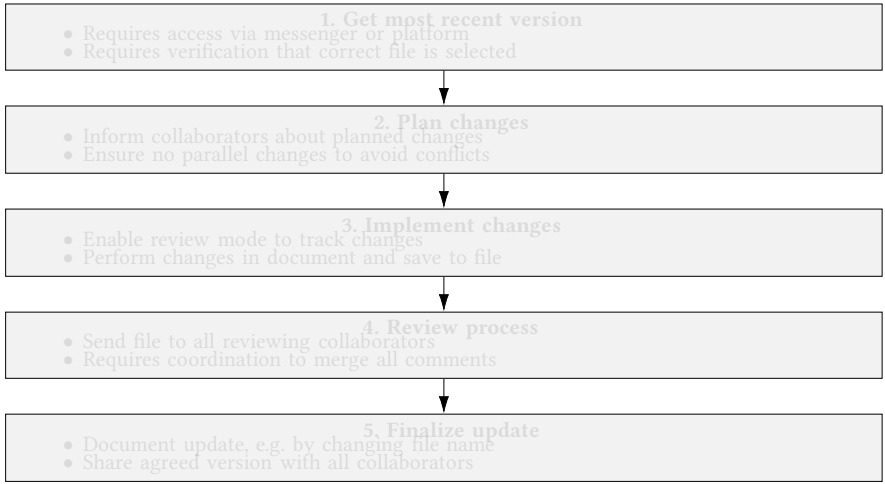

Collaborations on international standards for submission to \ac{ISO} such as \ac{MPEG} standards typically use the Word desktop applications.
Word files are exchanged per email or by hosting several file versions on a central NAS drive owned by the company.
Due to strict \ac{IP} company guidelines, the same process is commonly used for \ac{IP} relevant documents such as invention reports or patent applications.
Hence, online solutions such as \wftw are not allowed by the company guidelines.
To apply changes to an existing document, a collaborator usually follows the steps summarized in \autoref{fig:workflow_sota}.

"Who would work this way?", the attentive reader might ask.
These processes occur even in highly technical environments with well-educated employees.
There is significant potential for human error in this process, as humans need to undertake laborious tasks that can be automated.
Standardization efforts are a prominent example in which multiple companies with \ac{IP} right concerns collaborate on documents.
Their company policies often prohibit taking the risk of collaborating on shared document platforms with competitors, so the employees resort to the rather inefficient process of exchanging Word files.

\subsubsection{Research}

Researchers have a wide variety of backgrounds and significant liberties in designing their work environment.
Hence, they commonly apply varying techniques for collaborating on documents.
In research, the conventional workflow described in \autoref{sec:standards-workflow} is used as well as shared drives, \wftw, collaboration platforms such as Overleaf~\cite{overleaf_overleaf_2021}, or \magit.
A unified and efficient workflow could improve scientific collaborations and exchange of ideas.

\subsubsection{Software Development}

In professional software development, efficiency of the development process is critical for economic success.
Hence, companies scrutinize their tools and processes and strive for using efficient tools that enable their developers to work efficiently.
Successful companies rely heavily on using version control for their source code documents.
Most companies also use a variant of version control plus a markup language for documentation, such as Google's g3doc~\cite{winters_software_2020}, an internal wiki instance, or Confluence~\cite{atlassian_confluence_2021}.
Once more, this highlights that companies have found the combination of markup plus version control to be the most efficient collaborative documentation approach to date~\cite{winters_software_2020}.
We therefore next introduce an approach for efficient collaboration using a lightweight markup language and version control.

\section{Proposed Process}
\label{sec:process}

\begin{figure}[ht!]
    \centering
    \resizebox {.8\columnwidth} {!}{\scalefont{0.9}
\begin{tikzpicture}

    \tikzset{block/.style= {draw, rectangle, align=center,minimum width=1cm,minimum height=0.5cm}}

    \node[](anchor){};

    \node [block, text width = .75\columnwidth,fill=gray,fill opacity = 0.1, text opacity = 1, below = 0cm of anchor]  (get_repository) {
            \textbf{1. Get repository}
            \vspace{-.8em}
            \begin{itemize}
                \setlength\itemsep{-.4em}
                \item One repository with all documents
                \item Based on established git structures
            \end{itemize}};

    \node [block, text width = .75\columnwidth,fill=gray,fill opacity = 0.1, text opacity = 1, below = 0.5cm of get_repository]  (implement_changes) {
            \textbf{2. Implement changes}
            \vspace{-.8em}
            \begin{itemize}
                \setlength\itemsep{-.4em}
                \item Create new branch
                \item Perform changes and commit
            \end{itemize}};

    \node [block, text width = .75\columnwidth,fill=gray,fill opacity = 0.1, text opacity = 1, below = 0.5cm of implement_changes]  (finalize) {
            \textbf{3. Perform update}
            \vspace{-.8em}
            \begin{itemize}
                \setlength\itemsep{-.4em}
                \item Create pull request
                \item Merge after reviews of collaborators
            \end{itemize}};

    \path[draw, -{Latex[length=2.5mm,width=1.5mm]}]
    (get_repository.south) edge (implement_changes.north)
    (implement_changes.south) to (finalize.north);
\end{tikzpicture}}
    \caption{Overview of the proposed workflow for efficient large-scale collaborative writing.}
    \label{fig:workflow_proposed}
\end{figure}
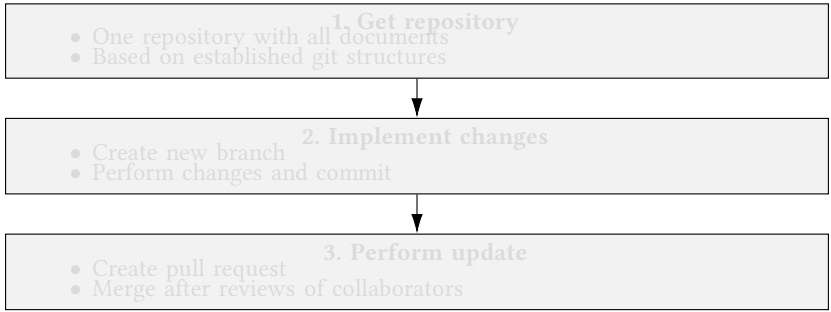

In this section, we present the proposed tooling and workflow for efficient collaborative writing.
The proposed documentation workflow is similar to workflows existing for collaboration on source code documents and incorporates many processes described in \autoref{sec:github}.
Each collaborator has a copy of the repository with all relevant documents.
When a collaborator wants to edit the document, a new branch for the changes is created.
Other collaborators can simultaneously work on the documents, as long as they have agreed to work on different sections/topics, for example through issues.
The entire process is summarized in \autoref{fig:workflow_proposed}.

While editing, the collaborator creates commits when atomic work-packages are completed.
When the work is done, the author creates a \ac{PR} in the shared repository, other collaborators review, and finally merge the changes.

We select the widely used \git~\cite{chacon_pro_2014} for a prototype of the proposed system.
\git enables the usage of collaboration platforms such as \gh or \gl.
These platforms are rather similar, in particular in terms of collaboration features.
We selected \gh because it has the larger market share of about \SI{88}{\percent}~\cite{gh_market_share_2020}.
The results of \cite{longo_use_2015} and \cite{marquardson_learning_2019} show that the proposed process contains significantly fewer manual, error-prone steps compared to \autoref{fig:workflow_sota}.
Note that the proposed approach is very straightforward since it combines established techniques and well known workflows from software development.
We argue that this is an important quality of the proposed workflow.

To further simplify the usability of the proposed approach, we provide a \gh template repository which allows to use the proposed workflow as a single-click solution.
The template repository can also be used with other platforms such as \gl.
It contains an AsciiDoc template and \ac{CI} configurations running automated checks and validation processes.
This template is publicly available on \gh\footnote{https://github.com/plain-docs/asciidoc-starter}.

\section{Evaluation}
\label{sec:comparison}

\ac{MPEG} is currently working on its documents using the workflow described in \autoref{sec:standards-workflow}.
We propose that \ac{MPEG} and similar bodies adopt the process proposed in \autoref{sec:process}.
This section compares the existing \ac{ISO} collaboration process of \autoref{sec:standards-workflow} based on the Microsoft Word desktop application, with the proposed collaboration workflow from \autoref{sec:process}.
Additionally, we compare both approaches with \wftw as the state-of-the-art approach for collaborative document creation.
In the following, we use "Word" to refer to both local Microsoft Word as well as \wftw.
For referring to one of the solutions, we explicitly name it.
We assess the aspects of accessibility, efficiency, traceability, interoperability, rendering, and error probability as well as the human factor.
An overview of the comparison is given in \autoref{tab:tldr}.

\begin{table*}[ht!]
    \centering
    \resizebox {\textwidth} {!} {
        \begin{tabular}{ L{0.15\textwidth} | C{0.3\textwidth} | C{0.3\textwidth} | C{0.3\textwidth}}
            \toprule
                                   & \textbf{Microsoft Word}                                                              & \textbf{\wftw}                                                                                 & \textbf{\magit}                                                                                 \\
            \midrule\midrule
            Primary output format  & Paper                                                                                & Paper                                                                                          & Web                                                                                             \\
            \midrule
            Editing mode           & Responsive \ac{WYSIWYG}                                                              & Responsive \ac{WYSIWYG}                                                                        & Syntax highlighting, live preview, any editor                                                   \\
            \midrule
            Sharing                & \red{Monolithic, local file}                                                         & \orange{Monolithic, shared file, available for all team members}                               & \green{Multiple files, \acp{URL} can link to anchors of documents}                              \\
            \midrule
            Scalable collaboration & \red{Limited, interfering reviews, single simultaneous edit}                         & \orange{Same reviews as Word, synchronous simultaneous edits, sufficient for small teams only} & \green{Proven, scalable review tools, (a-)synchronous simultaneous edits, separate discussions} \\
            \midrule
            Traceability           & \red{limited file history, multiple sources of truth}                                & \orange{file history, only automatic time-based versioning}                                    & \green{all advantages of \git, full traceability}                                               \\
            \midrule
            Interoperability       & \red{lock-in to proprietary files and software}                                      & \red{lock-in to proprietary files and software}                                                & \green{open source, extensible software, openness to other tools}                               \\
            \midrule
            Styling                & \red{Plethora of styling options leads to accidental issues/misuse}                  & \red{Similar to Word, but reduced set of options}                                              & \green{Few accidents through constrained, explicit styling. Rendering pipeline controls layout} \\
            \midrule
            Errors                 & \green{Powerful spelling and grammar checker}                                        & \green{Powerful spelling and grammar checker}                                                  & \green{Proven review process, option of additional verification tools}                          \\
            \midrule
            Setup                  & \green{Single tool}                                                                  & \green{Single tool}                                                                            & \red{Huge variety of tools might overwhelm novices}                                             \\
            \midrule
            Learning               & \green{Widely used among amateurs and professionals, GUI facilitates disoverability} & \green{Similar to Word}                                                                        & \orange{Known to professionals, but requires training for novices}                              \\
            \bottomrule
        \end{tabular}
    }
    \caption{Relevant features of using Microsoft Word, \wftw, and the proposed \magit process.}
    \label{tab:tldr}
\end{table*}

\subsection{Accessibility}

The way a document is displayed and shared defines its accessibility.
We discuss both accessibility aspects in this section.

\subsubsection{Document Display}
\label{sssec:document-consumption}

Over the past decades, the display of documents has changed fundamentally.
Several decades ago, the majority of documents still consisted of paper.
Since the invention of computers, digital documents have gained increasing importance for document exchange and processing.
Besides editing, sharing and processing digital documents on desktop computers, documents are nowadays also often used on mobile devices.
This change of medium also changes the requirements for displaying and processing the document.
When paper documents were prevalent, a document processing tool chain had to focus on creating documents laid out for printing to actual paper.
For many modern documents, paper is not the main medium anymore.
Instead, documents are viewed on screens of varying size and aspect ratio.
Consequently, modern document tool chains need to focus on these commonly web-based, diverse consumption scenarios.
A prominent example is Martin Flower's \textit{Refactoring: Improving the Design of Existing Code}~\cite{flower1999refactoring}, which is designed as a \textit{web-first} book.
Here, the online version of the book is the primary version, which contains more content than the physical book and the content is maintained over time.

Word is conventional in this respect.
Word documents mimic documents printed to physical paper, with a white background and defined document and font sizes.
These constraints limit accessibility of Word documents on web-based and mobile devices.

Markup languages, on the other hand, provide great support for modern, web-based document types.
Additionally, many formats such as AsciiDoc support publishing to classical paper as well as to digital formats.
Finally, plain text based files are future proof thanks to the simple file format.
The file itself can be used as an example of the format definition.
A prominent example for this is David Thomas' and Andrew Hunt's book \textit{The Pragmatic Programmer: your journey to mastery}~\cite{thomas2019pragmatic}, which is written in plain text and even recommends the benefits of using such.

\subsubsection{Document Sharing}

In an interconnected society in which shared resources are available to everyone, sharing is most efficiently done by not exchanging a file itself, but a pointer to the file.
In the dominant web-based environment, these file pointers are \acp{URL}.
Accessing an \ac{URL} in a web browser opens the desired document.
Word files can be shared and accessed this way, as can markup files.

Markup files have three significant advantages, however.
First, large documents can be split into multiple files, see \autoref{ssec:scalability}.
This allows straightforward, highly specific sharing.
If a user would like to share the entire document, this is simply done by sharing the \ac{URL} of the top-level file collecting all sub-files.
If sharing a specific section only, the user shares the \ac{URL} of that file only.

Second, markup languages such as HTML, Markdown, and AsciiDoc take that concept even further, allowing to create \acp{URL} to sections within one document.
This enables efficient sharing, as collaborators and readers can point to highly specific positions in their document with little effort.
Furthermore, the \acp{URL} are typically human readable which aids in communication.
To give one example, section \acp{URL} are widely used when sharing a specific style guide section of the extensive Google \Cpp Style Guide such as the style guide on static and global variables\footnote{https://google.github.io/styleguide/cppguide.html\#Static\_and\_Global\_Variables, Last accessed 09/26/2022}.
In markup source code, it is even possible to link to any line.

Third, direct accessibility is the core of markup languages.
A document consumer opens an \ac{URL}, and the web browser directly renders the document.
Word files take a different approach.
They require a dedicated application to render a Word file, requiring to use a \wftw instance or to download a document and open it in the dedicated application.
This contradicts modern internet usage, however, and often requires more time than a quick document lookup itself, disqualifying it in many applications.

\subsection{Efficiency \& Scalability}
\label{ssec:scalability}

This section focuses on the specifics of large documents and edits by large groups of collaborators.

\subsubsection{Text Editing}

First, a document processing system needs to provide an efficient text editing process.
WYSIWYG applications such as Word enable editing of the rendered output view.
This immediate feedback loop facilitates quick visual adjustments, but tends to become slow for large documents.

With plain text, one edits the source code of the document.
Widespread languages and text editors offer syntax highlighting for easy orientation inside the documents.
Given that plain text is simple to render, even long documents are rendered without delay.
Furthermore, many of the plain text editors allow advanced edits such as search and replace all, search with regular expressions, multiple cursors in the document, and more.
To merge the best from both worlds, many plain text formats offer live previews\footnote{https://docs.asciidoctor.org/asciidoctor/latest/tooling/\#visual-studio-code, Last accessed 09/26/2022}, updating the output e.g. each second given the set of source files.

\subsubsection{Collaboration}
\label{sssec:collaboration}

In the proposed process, a large document is structured into multiple files, which allows collaborators to effortlessly edit distinct files in parallel, without causing merge conflicts.
With Word's single \docx file, only one person at a time can edit the document.
With growing team sizes, this would quickly bring the editing process to a halt.
For comparison, imagine that only one of Google's engineers may work on its codebase at a time.
With \wftw, multiple collaborators can edit the same document simultaneously with live updates.

However, \wftw is still restricted to a single monolithic file.
A document split into multiple files simplifies navigation and allows to open only a subset of the document.
For large documents, both options become significant advantages compared to being forced to operate within one huge file.

Finally, the collaborative review process is facilitated by \gh.
Isolated discussions on different changesets (\acp{PR}) in the proposed process allow documents and collaborator teams to scale efficiently.
For example, at the time of this writing, there are 126 open \acp{PR} on the \Cpp standard~\cite{cpp_draft}, which is written in \LaTeX\xspace and hosted on \gh.
Each \ac{PR} has a distinct topic and discussion, without unnecessary interference between the discussions.
In contrast, there is no way of organizing all discussions taking place in the same Word document.
In Word, comments and suggestions provide a way of proposing changes without immediately changing the document.
While a comment always refers to a single place in the document, multiple comments across the document cannot be grouped into a single changeset, equivalent to a \ac{PR}.
This quickly leads to an unmanageable chaos as the number of simultaneous changes grows.

\subsection{Traceability}

A detailed, meaningful document history can be as important as the document itself.
Clear traceability of the document development timeline is important for collaborators~\cite{alwis_why_2009} and in a legal context~\cite{herkert2020boeing}.
It helps answering questions such as reasons for changes and responsibility.

First, we discuss versioning which is central to \acp{VCS} such as \git.
Significant effort has gone into making commits fast to create and apply alongside informative commit messages.
\git is the versioning system that dominates the software industry, hence it is expected to satisfy the vast majority of use cases.
The Word desktop application, on the other hand, just offers manual file saving.
In \wftw, the document is auto-saved regularly, without users being able to trigger saving.
Both options severely limit traceability compared to professional \acp{VCS}.

Second, the history of changeset discussions needs to be accessible in a persistent, structured way, which is the case with \gh \acp{PR}.
PRs are permanent recordings of an isolated discussion on a set of changes.
PRs can be labeled and searched for.
All requirements for discussion traceability are fulfilled.
Word is insufficient in this respect.
One can only search for discussions in the pool of all resolved and unresolved comments, while deleted comments are not visible anymore.
There is no option to search within the pool of comments.

\subsection{Interoperability \& Flexibility}
\label{ssec:interop}

Word locks projects into its ecosystem.
Transitioning to another program and process for collaborative document editing is difficult because of Word's proprietary, compressed file formats.
For many platforms such as Chrome \ac{OS} and Linux based devices, there is no native Word application.
Thus, users of these platforms have to resort to the web application, which has limited functionality as discussed in \autoref{ssec:word-online}.
With Word, a project depends on a commercial company's closed source software.
Users have heavily limited options for adapting the proprietary software to their needs.
Finally, users need to continuously pay software license fees.
All the above limitations do not exist for the \magit approach.
Plain text files can be easily parsed and modified with automated tools and can be edited with any editor. Given that the entire toolchain is open source, even significant case adaptions are feasible if sufficient resources can be invested.

Finally, plain text files parsed by automated tools enable new use cases.
Standard documents that contain syntax excerpts can have these syntax elements parsed by code generation.
The resulting code can subsequently be passed to compilers, which automatically uncover correctness or consistency issues.
This can be done continuously in automated tasks that are triggered in the \gh or \gl repository using \ac{CI} tools.

As a more practical example, Hofbauer~et~al.~\cite{hofbauer_software_lab} used this approach for creating a lecture in which the lecture slides are written in Markdown.
The lecture slides contain descriptions of the homework tasks for the students.
Using comment-based delimiters not visible on the lecture slides, the authors separate a task from the remaining slide content.
This allowed them to build a simple toolchain that can create issues for all student groups with the task title and description from the slides with a single command.
Such an approach scales well and enables a single definition of each homework task.
This would not be possible with Microsoft PowerPoint-based lecture slides.
These are just two examples for what is possible when document content is easily accessible from other tools.
If plain text formats become the norm, we expect many more such processes to be invented.

\subsection{Output Styling}

In the \ac{ISO} standardization process, collaborators frequently change style by accident, e.g. to a slightly different font or font size.
This happens because style is implicit in Word and because an abundance of styling options is offered.
Such changes leads to inconsistent documents, requiring manual fixing effort.
Usually a single \ac{ISO} contributor spends several hours or even days in creating a consistent layout when a submission is due.

In markup toolchains, styling options are constrained and a deviation from the consistent style is explicitly instructed.
These restrictions lead to fewer accidents and hence higher quality documents and less manual effort.
Style selection such as the rendering font is a question for the rendering tool, not for the author while writing the text.
Markup also allows for consistent integration into a particularly styled environment without the document imposing hard requirements on e.g. font, font size, and page layout.

\subsection{Errors}

Word supports authors with sophisticated spell and grammar checkers, preventing many language issues.
However, technical and higher level errors still can only be prevented through human review.
While Word provides a side by side comparison of changes and inline suggestions, this lacks behind the \git review process, as presented in \autoref{sssec:collaboration}.
Further, the plain text files of markup languages can be automatically validated for technical errors.
Such checks can be executed locally during the creation process and in \ac{CI} as an additional check before the changeset is integrated.
As mentioned in \autoref{ssec:interop}, the sky is the limit for what people invent.

\subsection{Generalizability}

The benefits outlined in this section, summarized in \autoref{tab:tldr}, are not restricted to the \ac{ISO} standardization process, but applicable to all processes in which a group of contributors has to work on a common document.

While the benefits of \git and \gh are not new, many processes such as \ac{ISO} standardization are still done working on a single Word document and can be optimized.
The proposed approach offers a unified, flexible, and modern way for creating such documents relying on established and well know workflows of the software development domain.

\subsection{Human Factor}

To maximize the benefits of the proposed \magit workflow, users require more education to handle the higher technical complexity compared to the basic Word workflow.
Since we often referred to the \ac{ISO} standardization process as an example, we can expect contributors with a solid technical background where most of them already use tools related to \magit in their daily workflow.
For this user group, using the same process for document creation should be seamless.
Even novice users with a less technical background can learn these processes.
The available platforms such as \gh or \gl already provide web IDEs to make the first steps for novice users as easy as possible.

\section{Conclusion}
\label{sec:conclusion}

In this paper, we presented a flexible and scalable approach for collaborative creation of documentation or specifications in large teams.
While the majority of specifications, such as for \ac{ISO} standardization, are still created in a single, large Word document, we propose to use well-established processes from software development together with a lightweight markup language.
We performed a thorough comparison to the most widespread workflow using Microsoft Word and discussed the benefits and drawbacks of the proposed approach.
The proposed approach comes with several benefits such as plain text based document sources that allow for higher levels of automated validation processes.
Another major benefit is the collaboration and review support provided by the well-established tool \git and collaboration platforms such as \gh or \gl, which are specifically designed for collaborative creation.
The higher technical complexity of the proposed \magit approach requires users with more education compared to the basic Word workflow.

We provide a template repository publicly available on \gh that can be used as a one-click solution to set up a collaborative document creation process such as required for the \ac{ISO} standardization.

For future work, we plan to further assess the proposed approach using objective metrics.
Conducting an empirical study of a large-scale projects such as the \Cpp \ac{ISO} draft~\cite{cpp_draft} can yield valuable quantitative insights into the benefits of the proposed \magit approach for efficient collaborative writing.

\bibliographystyle{ACM-Reference-Format}
\bibliography{references}

\end{document}